\documentclass[aps,prb,twocolumn,groupedaddress,superscriptaddress,nofootinbib]{revtex4}
\usepackage{graphicx}

\begin{document}

\title{Application of metasurface description for multilayered metamaterials and an alternative theory for metamaterial perfect absorber}

\author{Jiangfeng Zhou$^{*\dag}$}
\affiliation{Center for Integrated Nanotechnologies, Los Alamos National Laboratory, Los Alamos, New Mexico 87545, USA}

\author{Hou-Tong Chen$^*$}
\affiliation{Center for Integrated Nanotechnologies, Los Alamos National Laboratory, Los Alamos, New Mexico 87545, USA}

\author{Thomas Koschny}
\affiliation{Ames Laboratory and Department of Physics and Astronomy, Iowa State University, Ames, Iowa 50011, USA}

\author{Abul K. Azad}
\affiliation{Center for Integrated Nanotechnologies, Los Alamos National Laboratory, Los Alamos, New Mexico 87545, USA}

\author{Antoinette J. Taylor}
\affiliation{Center for Integrated Nanotechnologies, Los Alamos National Laboratory, Los Alamos, New Mexico 87545, USA}

\author{Costas M. Soukoulis}
\affiliation{Ames Laboratory and Department of Physics and Astronomy, Iowa State University, Ames, Iowa 50011, USA}

\author{John F. O'Hara$^{*\ddag}$}
\affiliation{Center for Integrated Nanotechnologies, Los Alamos National Laboratory, Los Alamos, New Mexico 87545, USA}

\date{\today}
\begin{abstract}
We analyze single and multilayered metamaterials by modeling each layer as a metasurface with effective surface electric and magnetic susceptibility derived through a thin film approximation. Employing a transfer matrix method, these metasurfaces can be assembled into multilayered metamaterials to realize certain functionalities. We demonstrate numerically that this approach provides an alternative interpretation of metamaterial-based perfect absorption, showing that the underlying mechanism is a modified Fabry-Perot resonance. This method provides a general approach applicable for decoupled or weakly coupled multilayered metamaterials.

\end{abstract}


\maketitle

\let\thefootnote\relax\footnotetext{$^*$Correspondence should be addressed to J. F. Zhou (jfengz@gmail.com), H.-T. Chen (chenht@lanl.gov) or J. F. O'Hara (oharaj@okstate.edu)\\
$^\dag$Current address: Department of Physics, University of South Florida\\
$^\ddag$Current address: Department of Electrical and Computer Engineering, Oklahoma State University}
Electromagnetic (EM) metamaterials are artificial materials that can be engineered to exhibit controlled optical properties not found in nature
over most of the EM spectrum \cite{Sci_soukoulis_2007,Nat_Phonotics_shalaev_2007}.  They usually consist of multiple identical layers of periodically arranged artificial structures, and are considered bulk homogeneous media with constitutive parameters obtained by a retrieval based on effective medium theory \cite{PRB_smith_retrieval_2002,PRL_koschny_2004}. Recently, heterogeneous metamaterials consisting of two or more \emph{distinct} layers were used to realize functionalities such as perfect absorption at THz \cite{Perfect_absorber_Padilla_PRL_2008} and infrared frequencies \cite{Perfect_absorber_Padilla_PRL_2010,Perfect_absorber_Giessen_Nano_Lett_2010,Zhou_Lei_APL_Aborober_2010}, and EM wave tunneling \cite{EM_Tunneling_PRL_Zhou_Lei_2005}. In that work, each layer of the metamaterial \cite{EM_Tunneling_PRL_Zhou_Lei_2005} or all the layers as an entirety \cite{Perfect_absorber_Padilla_PRL_2008,Perfect_absorber_Padilla_PRL_2010,Perfect_absorber_Giessen_Nano_Lett_2010} were considered as a homogeneous medium. The effective permittivity and permeability, were calculated using an established retrieval procedure \cite{PRB_smith_retrieval_2002,PRL_koschny_2004}. However, the metamaterials in these systems consist of only \emph{one} functional layer of artificial structures (meta-atoms), which is analogous to a single molecular layer in natural materials. It is challenging to define bulk effective permittivity and permeability for such single-``meta-atom"-layer systems, since these macroscopic material properties typically result from averaging fields over many molecular layers. In addition the thickness of the effective bulk material in these systems is not uniquely defined, which therefore likewise renders the effective material properties arbitrary \cite{zhou_photon_nano_2008,Holloway_metafilm_2009}. Further complications arise because metamaterials consisting of \emph{distinct} layers, such as perfect absorbers \cite{Perfect_absorber_Padilla_PRL_2008,Perfect_absorber_Padilla_PRL_2010,Perfect_absorber_Giessen_Nano_Lett_2010}, are inhomogeneous in the wave propagation direction, and cannot be strictly considered homogeneous bulk media.

In this paper, we use an effective medium model that treats each layer of the metamaterial as a metasurface with unique effective surface electric and magnetic susceptibility, $\chi_{se}$ and $\chi_{sm}$. Through a thin film approximation, we obtain the same equations of $\chi_{se}$ and $\chi_{sm}$ as previous metasurface work \cite{Holloway_metafilm_2009}, and also reveal the relations between surface effective susceptibilities and bulk effective material parameters. We then use a transfer matrix method to analyze the overall EM properties of multilayered metamaterials using the effective material parameters (surface susceptibilities) of each layer. We find that the overall properties of multilayered metamaterials can be determined by their individual layer properties, in the absence of inter-layer resonance coupling. We also find that such individual layer properties are responsible for metamaterial perfect absorbers. This contrasts with previous explanations based on bulk effective medium theory \cite{Perfect_absorber_Padilla_PRL_2008,Perfect_absorber_Padilla_PRL_2010,Perfect_absorber_Giessen_Nano_Lett_2010}. To wit, in previous work, the \emph{entire} metamaterial was considered as a homogeneous medium with independently engineered effective permittivity and permeability to reach the condition $\epsilon_{\mathrm{eff}}=\mu_{\mathrm{eff}}$, both having large imaginary part resulting in the effective refractive index, $n=n^{'}+\mathrm{i}n^{''}$ and $n^{''}\gg n^{'}$. The EM wave thus propagates through the first interface (air-metamaterial) without reflection and the strength decays rapidly to zero inside the metamaterial before reaching the second interface (metamaterial-air). However, our results show that the interaction (or the assumed magnetic resonance) between the two metallic layers has a negligible effect on the absorption. Instead, the functional mechanism is the Fabry-Perot interference resulting from the multiple reflections in the cavity bounded by two metamaterial layers. Finally, we also find that the metamaterial EM tunneling \cite{EM_Tunneling_PRL_Zhou_Lei_2005} and the metamaterial anti-reflection \cite{Chen_anti_reflection_PRL_2010} can be explained very well by our approach. Our approach is generally applicable for decoupled or weakly coupled multilayered metamaterials \cite{PRB_zhou_strongly_weakly_coupled_MM}, where the coupling due to evanescent modes is inconsequential.
%
\begin{figure}[htb]\centering
  \includegraphics[width=7cm]{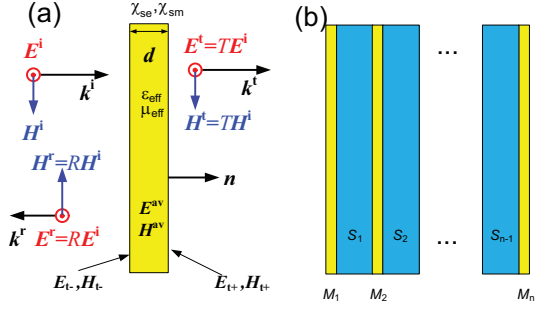}\\
  \caption{(a) A schematic of a single layer metamaterial considered as a homogeneous thin film and the electric and magnetic field across it. $\textbf{E}^{\mathrm{i}}$, $\textbf{E}^{\mathrm{r}}$ and $\textbf{E}^{\mathrm{t}}$ represent transverse electric field of the incident, reflected and transmitted EM wave under normal incidence, respectively; $\textbf{E}_{t-}$, $\textbf{E}_{t+}$, $\textbf{H}_{t-}$ and $\textbf{H}_{t+}$ are the total transverse electric and magnetic fields at each boundary of the film; $\textbf{E}^{av}_{t}$ and $\textbf{H}^{av}_{t}$ are the average transverse electric and magnetic fields inside the film; $\chi_{\mathrm{se}}$ and $\chi_{\mathrm{sm}}$ represent the effective surface electric and magnetic susceptibilities; $\epsilon_{\mathrm{eff}}$ and $\mu_{\mathrm{eff}}$ represent the effective permittivity and permeability. (b) Schematic of a multilayered metamaterial consisting of N layers separated by N-1 layers of dielectric spacers.
  \label{fig_geom}}
\end{figure}

We begin with Fig. 1(a), where a single-layer metamaterial is considered as a homogeneous thin film with thickness, $d$, same as the actual thickness of a single-layered of metamaterial structure and approaching zero as compared to the incident wavelength. Transmission and reflection occurs as a plane EM wave propagates normally through the thin film, and generally leads to discontinuities of the transverse electric and magnetic fields, which can be described by the following boundary conditions \cite{tretyakov_EM_book}:
\begin{eqnarray}\label{eqn_BC}
\textbf{n}\times(\textbf{E}_{t+}-\textbf{E}_{t-})&=&\mathrm{i}\omega\mu_0\mu_{\mathrm{eff}} d\textbf{H}^{av}_{t}\\
\textbf{n}\times(\textbf{H}_{t+}-\textbf{H}_{t-})&=&-\mathrm{i}\omega\epsilon_0\epsilon_{\mathrm{eff}} d\textbf{E}^{av}_{t}
\end{eqnarray}
where $\textbf{E}_{t-}=(1+R)\textbf{E}^\mathrm{i}$, $\textbf{E}_{t+}=T\textbf{E}^\mathrm{i}$, $\textbf{H}_{t-}=(1-R)\textbf{H}^\mathrm{i}$, $\textbf{H}_{t+}=T\textbf{H}^\mathrm{i}$, $\textbf{E}^{av}_{t}$ and $\textbf{H}^{av}_{t}$ are defined in the Fig. 1(a) caption. The average electric and magnetic fields inside the thin film, can be approximately defined as $\textbf{E}^{av}_{t}=(\textbf{E}_{t-}+\textbf{E}_{t+})/2=(1+R+T)\textbf{E}^\mathrm{i}/2$ and $\textbf{H}^{av}_{t}=(\textbf{H}_{t-}+\textbf{H}_{t+})/2=(1-R+T)\textbf{H}^\mathrm{i}/2$ for very thin films, i.e., $d\ll\lambda$. $\epsilon_{\mathrm{eff}}$ and $\mu_{\mathrm{eff}}$ are the effective permittivity and permeability of the thin film. The right-hand side of equations (1) and (2) contains the bulk magnetic and electric current densities, $J_m=-\mathrm{i}\omega\mu_{\mathrm{eff}}\textbf{H}^{av}_{t}$ and $J_e=-\mathrm{i}\omega\epsilon_{\mathrm{eff}}\textbf{E}^{av}_{t}$. In the limit $d\ll \lambda$, the thin film can also be equivalently considered as a single interface (metasurface) with surface current density, $J_{se}=\int J_e \mathrm{d}z=J_ed$ and $J_{sm}=\int J_m \mathrm{d}z=J_md$, resulting from the discontinuity of transverse electric and magnetic fields across the thin film, respectively. The surface electric and magnetic current densities can be characterized by effective surface electric and magnetic susceptibilities, $J_{se}=-i\omega\epsilon_0\chi_{se}\textbf{E}^{av}_t$ and $J_{sm}=-i\omega\mu_0\chi_{sm}\textbf{H}^{av}_t$. We can obtain $\chi_{se}=(\epsilon_{\mathrm{eff}}-1)d$, $\chi_{sm}=(\mu_{\mathrm{eff}}-1)d$, where the constant $1$ results from the permittivity or permeability of vacuum when replacing a finite thickness slab by a zero thickness surface. Using the previous equations for average fields, $\chi_{se}$ and $\chi_{sm}$ can now be expressed as functions of the complex transmission and reflection coefficients $T$ and $R$:
\begin{eqnarray}\label{eqn_BC}
\chi_{se}&=&\frac{2\mathrm{i}}{k_0}\frac{1-R-T}{1+R+T}\\
\chi_{sm}&=&\frac{2\mathrm{i}}{k_0}\frac{1+R-T}{1-R+T}
\end{eqnarray}
where $k_0$ is the wavevector in vacuum. Equations (3) and (4) are in consistent with previous work \cite{Holloway_metafilm_2009}, except for a sign reversal for $\chi_{sm}$ in Ref. 10, which we believe is a misprint.

Since the transmission and reflection coefficients are independent from the thickness, $d$, the surface susceptibilities, $\chi_{se}$ and $\chi_{sm}$, are well-defined parameters describing the properties of single-layered metamaterial in isolation. Hence they are distinct from the effective parameters of bulk metamaterials, $\epsilon_{\mathrm{eff}}$ and $\mu_{\mathrm{eff}}$, which are non-unique and depend on the effective metamaterial thickness $d$. Despite this, we also find that the effective permittivity and permeability of individual metamaterial layers calculated using a usual retrieval procedure \cite{PRB_smith_retrieval_2002} show good consistency with $\epsilon_{\mathrm{eff}}=\chi_{se}/d+1$, and $\mu_{\mathrm{eff}}=\chi_{sm}/d+1$ obtained from equations (3) and (4). The main exception is the anti-resonance artifacts obtained from the retrieval procedure and due to periodicity effect are absent in the effective surface susceptibilities. This means common retrieval procedures may be used to obtain the surface susceptibilities of single-layer metamaterials with some accuracy.

Employing a transfer matrix method, we can determine the overall transmission and reflection of a decoupled multilayered metamaterial from the effective material parameters derived for each layer. In the following, we use a metamaterial perfect absorber in Ref. \cite{Perfect_absorber_Padilla_PRL_2010} as an example to demonstrate how to apply this approach and reveal its underlying mechanism.

%
Figure 2(a) shows a perfect absorber metamaterial \cite{Perfect_absorber_Padilla_PRL_2010} consisting of two layers of metallic structures separated by a dielectric spacer. The first metallic structure is an array of cross-wire resonators and the second is a metallic ground plane. Each can be modeled as a metasurface respectively. The whole metamaterial is then considered as a three-layered system consisting of two metasurfaces separated by a dielectric spacer.

In the perfect absorber metamaterial, the cross-wire structure exhibits electric resonance modes with resonance frequencies determined by its structural parameters. To obtain the effective material parameters of metamaterials, we performed numerical simulations with CST Microwave Studio (Computer Simulation Technology GmbH, Darmstadt, Germany), which uses a finite-difference time-domain method to determine $R$ and $T$ of the metallo-dielectric structures. The unit cell used in the simulation for the first layer, $\mathrm{MM}_1$, is schematically shown as the inset in Fig. 2(b). It consists of a gold cross-wire with thickness, $d_1=0.1\ \mu m$, width $w=0.4\ \mu m$, length $l=1.7\ \mu m$ and period $a=2\ \mu m$, on the spacer layer with thickness, $s=0.185\ \mu m$, and dielectric constant, $\epsilon_s=2.28(1+0.04\mathrm{i})$. Gold is modeled as a Drude metal with a plasma frequency, $f_p=2181$ THz, and damping frequency, $f_\tau=6.5$ THz \cite{Novel_metal_property_Ordal_Applied_Opt_1985}. Then the effective material parameters of the metamaterial, $\mathrm{MM}_1$, bounded by vacuum, are calculated using equations (3) and (4) with slight modification to handle the substrate surrounding the metamaterial structure \cite{Zhao_Chiral_sub_retrieval_Opex_2010}. Similarly, the gold plate with thickness, $d_2=0.2\hspace{1 mm}\mu m$, was modeled as the second layer using a unit cell shown as $\mathrm{MM}_2$ in the inset of Fig. 2(b). The calculated effective surface electric susceptibility of the metamaterial is shown in Fig. 2(b), where the cross-wires exhibit electric resonances at wavelengths of 4.85 $\mu m$ and 1.68 $\mu m$, and the gold plate exhibits a plasmonic response with large negative permittivity over the entire wavelength range from 1.5 to 8 $\mu m$. Importantly, the effective magnetic susceptibility (not shown here) for $\mathrm{MM}_1$ and $\mathrm{MM}_2$ was calculated to be a constant close to zero over the entire wavelength range. Figure 2(c) shows two absorption peaks at the wavelength of 1.86 and 6.18 $\mu m$, obtained by full EM simulations of the entire multilayered metamaterial (solid curve). The latter peak corresponds to the absorption reported in Ref. \cite{Perfect_absorber_Padilla_PRL_2010}.
%
\begin{figure}[htb]\centering
  \includegraphics[width=6.5cm]{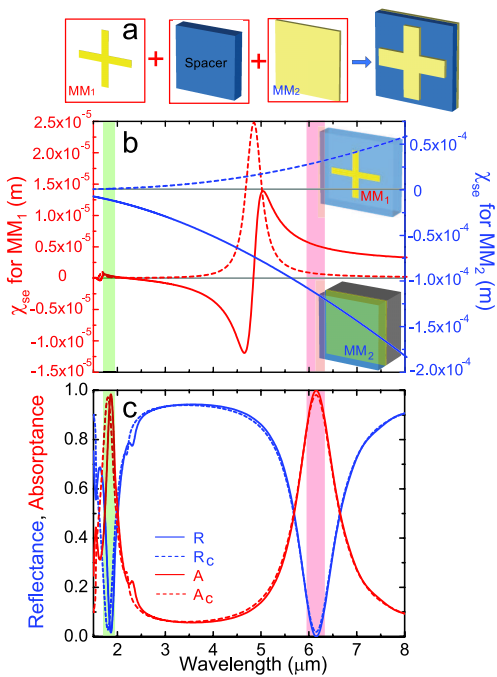}\\
  \caption{(a) The perfect absorber metamaterial is modeled as a stack of three layers, the cross-wire metamaterial, a dielectric spacer, and the metallic plate. (b) The real (solids curves) and imaginary (dashed curves) parts of the effective surface electric susceptibility of the metasurface representing cross-wires (red) and metallic plate (blue). The inset $\mathrm{MM}_1$ and $\mathrm{MM}_2$ shows the unit cells used in numerical simulations to obtain the metasurface parameters. (c) The solid curves show the reflectance (blue), $R$, and absorptance (red), $A$, obtained from direct simulations, while the dashed curves show the calculated values, $R_c$ (blue), $A_c$ (red), using a 3-layer metamaterial model.
  \label{fig_epsilon_z}}
\end{figure}
To determine the behavior of the whole absorber we first derive the transfer matrix of the individual layers. The transfer matrix of a metasurface, bounded by vacuum on each side, can be determined from the relation between the transfer matrix and S-parameter matrix,


\begin{equation}\label{eqn_T_S}
\hspace{-0.15in}M=\left(\matrix{%
M_{11} & M_{12}\cr%
M_{21} & M_{22}}\right)
=\left(\matrix{%
S_{12}-S_{11}S_{22}S_{21}^{-1} & S_{11}S_{21}^{-1}\cr%
-S_{21}^{-1}S_{22} & S_{21}^{-1}}\right)
\end{equation}
where $S_{21}$, $S_{12}$ are forward and backward transmission coefficients, and $S_{11}$, $S_{22}$ are reflection coefficients at front and back sides of the metasurface, respectively. Equation (5) also applies to the dielectric spacer. All of the individual transfer matrices are now multiplied to obtain the total transfer matrix of the whole metamaterial, $M^{\mathrm{tot}}$=$M_{\mathrm{MM}_1}M_{\mathrm{S}}M_{\mathrm{MM}_2}$. This now constitutes a full description of the metamaterial perfect absorber based on the effective parameters of the individual metasurfaces. Using the relation between the transfer matrix and S-parameter matrix again, we can obtain the transmission and reflection coefficients of the whole metamaterial, $\widetilde{T}=S_{21}=1/M_{22}^{\mathrm{tot}}$ and $\widetilde{R}=S_{11}=M_{12}^{\mathrm{tot}}/M_{22}^{\mathrm{tot}}$. As shown in Fig. 2(c), the calculated reflectance, $R_c=|\widetilde{R}|^2$, and absorptance, $A_c=1-|\widetilde{R}|^2-|\widetilde{T}|^2$, agree very well with the corresponding $R$ and $A$ obtained from a direct simulation of the whole metamaterial. Since the transfer matrix calculations only take account of the transmissions and reflections between individual layers, and since each layer's properties were determined in isolation, this shows that any inter-layer resonance coupling occurring between the cross-wire and the metallic plate layers is inconsequential. Hence the magnetic response reported in previous absorber work \cite{Perfect_absorber_Padilla_PRL_2008,Perfect_absorber_Padilla_PRL_2010,Perfect_absorber_Giessen_Nano_Lett_2010}, is unlikely to have any significant functional effect, since it relies on strong coupling between two metallic layers in the form of anti-parallel resonance currents. To further understand the magnetic response, we examined the double-fishnet structure \cite{science_soukoulis_2006}, where the magnetic resonance mode exists due to the strong coupling between two metallic layers. As we expected, the transfer matrix calculation failed to reproduce the magnetic resonance in the direct simulations of whole double-fishnet structure since it violates the decoupling or weakly coupling assumption.

To better understand the mechanism of the perfect absorber, we derived the overall reflection coefficient, $\widetilde{R}$, using the mathematical form of a slightly modified Fabry-Perot cavity, in terms of the transmission and reflection coefficients of metasurfaces:
\begin{equation}\label{eqn_R}
\widetilde{R}=\frac{R_{12}+\alpha R_{23}e^{2\mathrm{i}\beta}}{1-R_{21}R_{23}e^{2\mathrm{i}\beta}}
\end{equation}
where, $T_{21}$, $T_{12}$, $R_{12}$ and $R_{21}$, are transmission and reflection coefficients of the metasurfaces, $\mathrm{MM}_1$, regarded as an interface bounded by semi-infinite media. They are generally functions of effective surface electric and magnetic susceptibilities of $\mathrm{MM}_1$. They can also be obtained by numerical simulation of $\mathrm{MM}_1$ using the structure shown in Fig. 2(b). $R_{23}=-1$ is the reflection coefficients from the gold ground plane, $\mathrm{MM}_2$; $\beta=n_skd_s$ is the propagating phase term in the spacer; and $\alpha=T_{21}T_{12}-R_{12}R_{21}$. At the perfect absorbing wavelength, the reflection coefficient, $\widetilde{R}=0$, which requires the following conditions:
\begin{eqnarray}
&&|R_{12}|=|\alpha|\label{eqn_PA1}\\%
&&\phi(R_{12})-\phi(\alpha)-2\beta=2m\pi, \quad|m|=0,1,2,... \label{eqn_PA2}
\end{eqnarray}
%
\begin{figure}[htb]\centering
  \includegraphics[width=6.5cm]{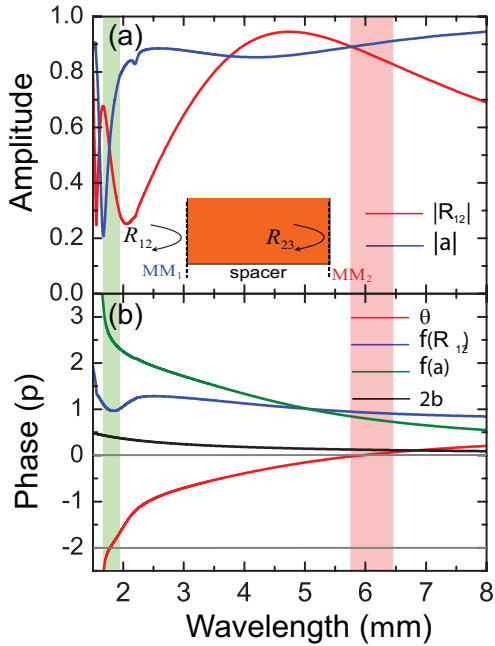}\\
  \caption{(a) The amplitude of $R_{12}$ (blue) and $\alpha$ (red). (b) The phase terms in equation (\ref{eqn_PA2}), $\theta=\phi(R_{12})-\phi(\alpha)-2\beta$ (red), the phase of $R_{12}$ (blue) and $\alpha$ (green), and the propagation phase $2\beta$ (black) are shown respectively.
  \label{fig_AR_condition}}
\end{figure}
To understand these conditions, similar to recent work on anti-reflection metamaterials \cite{Chen_anti_reflection_PRL_2010}, we calculated the amplitude and phase terms shown in equations (\ref{eqn_PA1}) and (\ref{eqn_PA2}) using the effective material parameters of the metasurfaces.  As shown in Fig. 3(a), in the strongly absorbing regions (shaded) centered at the wavelengths of 6.18 and 1.86 $\mu m$, the amplitudes of $R_{12}$ and $\alpha$ are almost equal, roughly satisfying the amplitude condition. In Fig. 3(b), the phase term, $\theta$ crosses zero and 2$\pi$ at wavelengths of 6.18 and 1.86 $\mu m$, respectively, indicating the phase condition in equation (\ref{eqn_PA2}) is perfectly fulfilled at the absorption peaks. Several other absorption peaks (not shown here) can be observed at shorter wavelengths when $\theta$ reaches 4$\pi$, 6$\pi$ etc. Figure 3 also shows that equations (\ref{eqn_PA1}) and (\ref{eqn_PA2}) are not simultaneously satisfied at the same wavelengths, which explains why the reflectance $|\widetilde{R}|^2$ (blue dashed curve in Fig. 2(b)) does not reach zero. Equations (\ref{eqn_R}-\ref{eqn_PA2}) and Fig. 3 indicate that the nature of the absorber is Fabry-Perot-like resonance modes resulting from multiple wave reflections between metasurfaces $\mathrm{MM}_1$ and $\mathrm{MM}_2$. The metasurface $\mathrm{MM}_1$ and ground plane $\mathrm{MM}_2$ form a Fabry-Perot cavity filled with a lossy dielectric spacer. Strong absorption occurs as the EM wave propagates through the lossy dielectric spacer multiple times. The metasurface $\mathrm{MM}_1$ provides the proper surface susceptibility to fulfill the conditions presented in equations (\ref{eqn_PA1}) and (\ref{eqn_PA2}). Practically, the understanding of this mechanism helps us to improve the metamaterial designs. For instance, to achieve a perfect absorption, we need to optimize the design of the metamaterial, $\mathrm{MM}_1$, to reach the conditions in equation (\ref{eqn_PA1}) and (\ref{eqn_PA2}) simultaneously. Adjusting the slope of $|R_{12}|$, $|\alpha|$ and $\theta$  curves shown in Fig. 3 can also optimize the absorbing bandwidth of the whole metamaterial.

In conclusion, we have proposed an effective medium model for individual metasurfaces. In our model, each layer of the multilayered metamaterial is considered as a metasurface. These metasurfaces may be combined to determine the properties of multilayered metamaterials using the transfer matrix method. This alternate interpretation resolves the problems of defining a multilayered metamaterial as a single-``meta-atom"-layered bulk medium. This method provides a general approach applicable for any decoupled or weakly coupled multilayered metamaterials. We applied this method to the recently demonstrated perfect absorber metamaterials and identified the underlying mechanism as Fabry-Perot type resonance modes in contrast to the previously reported mechanism of independent engineering of the bulk effective permittivity and permeability. We have also found that this model accurately reproduces the previously reported EM wave tunneling effects \cite{EM_Tunneling_PRL_Zhou_Lei_2005}.

We acknowledge support from the Los Alamos National Laboratory LDRD Program. This work was performed, in part, at the Center for Integrated Nanotechnologies, a US Department of Energy, Office of Basic Energy Sciences Nanoscale Science Research Center operated jointly by Los Alamos and Sandia National Laboratories. Work at Ames Laboratory was supported by the Department of Energy (Basic Energy Sciences) under contract No. DE-AC02-07CH11358. This was partially supported by the U.S. Office of Naval Research, Award No. N000141010925. We thank Christopher Holloway, Willie Padilla and Richard Averitt for helpful discussions.
%
%
%

\end{document}